\documentclass{aa}
\usepackage{graphicx}
\usepackage{graphicx,subcaption}
\usepackage{txfonts}
\usepackage{natbib}
\usepackage[bookmarksopen=false,bookmarksnumbered=true,breaklinks=true,colorlinks=true,linkcolor=blue,citecolor=blue]{hyperref}

\usepackage{multicol}
\usepackage{soul} 
\setstcolor{red}
\newdimen\captwidth
\newdimen\figwidth
\newcommand{\mjup}{\ensuremath{\textnormal{M}_{\textnormal{Jup}}}}

\usepackage{soul}
\usepackage{xcolor}

\begin{document}
\title{Dynamics of the TWA~7 planetary system and possibility of an additional planet}

\author{A. Lacquement\inst{1} \and  H. Beust\inst{1} \and G. Duch\^ene\inst{1,2} \and A.-M. Lagrange\inst{1,3}}

\institute{
$^{1}$Univ. Grenoble Alpes, CNRS, IPAG, 38000 Grenoble, France\\
$^{2}$Department of Astronomy, University of California, Berkeley, CA 94720, USA\\
$^{3}$LIRA, CNRS, Observatoire de Paris, Université PSL, 92190 Meudon, France}
\date{Received 26 January 2026 / Accepted 28 February 2026}  \offprints{A. Lacquement}
\mail{antoine.lacquement@univ-grenoble-alpes.fr}
\titlerunning{Dynamics of the TWA~7 planetary system}
\authorrunning{A. Lacquement et al.}
\abstract
{
The debris disk surrounding the young star TWA~7 exhibits morphological features that tightly constrain its planetary architecture. JWST/MIRI observations have recently revealed a directly imaged outer planet orbiting at a large separation. The disk also displays a sharply defined inner edge at $\sim$23~au and an extended asymmetric structure that may trace a horseshoe-like distribution of material indicative of gravitational interactions between planets and planetesimals.
}
{
We investigate whether the observed disk morphology and the possible presence of co-orbital material can be explained by the combined gravitational influence of the known outer planet and an undetected inner companion. We aim to identify planetary configurations consistent with both the disk structure and the long-term dynamical stability of the system.  
}
{
We combined N-body simulations and secular perturbation theory to explore how an undetected inner planet could shape the inner edge of the disk while maintaining the dynamical coldness required for stable co-orbital structures around the outer planet. The analytical framework quantifies the secular coupling between the two planets and delineates dynamically viable configurations.
}
{
The inner edge of the disk at $\sim$23~au can be reproduced by a sub-Jovian planet orbiting between 13 and 23~au. Secular interactions further restrict this companion to nearly circular orbits, as higher eccentricities would excite the outer planet and destabilize the co-orbital material. Together, these constraints confine the system to a narrow region of parameter space.   
} 
{
The TWA~7 system appears dynamically cold, with all components, including the planets and the debris disk, sharing nearly circular and coplanar orbits. Such a quiescent configuration likely reflects the weak dynamical stirring, making it a promising laboratory to study the early interplay between planet formation, co-orbital dynamics, and debris-disk evolution.
}

\maketitle
\section{Introduction}
\label{introduction}

Debris disks, as remnants of planet formation, offer crucial insights into the architecture and dynamical evolution of young planetary systems \citep[e.g.,][]{Wyatt+2008, Hughes+2018}. Their observed morphologies, such as gaps, warps, or sharp edges, often trace the gravitational imprint of unseen planets. A notable example is the $\beta$~Pictoris system, whose warped disk was long interpreted as indirect evidence of a giant planet \citep{Mouillet+1997}, which was later confirmed by direct imaging \citep{Lagrange+2009}. Such cases illustrate how disk structures can guide the detection and characterization of planets, while dynamical modeling helps identify additional undetected companions and improves understanding of their evolution history.

The nearby M-type star TWA~7 lies at a distance of $\sim$34~pc \citep{Bailer-Jones+2018} and has a mass of $\sim$0.46~$M_{\odot}$ \citep{Lagrange+2025}. As a member of the TW~Hydrae association, its age is estimated to be $\sim$10~Myr \citep{BarradoYNavascues+2006, Bell+2015}. Recently, observations with the James Webb Space Telescope (JWST) using the Mid-Infrared Instrument (MIRI) led to the direct detection of a planet, TWA~7~b, at a projected separation of $\sim$52~au \citep{Lagrange+2025}. Its infrared photometry, interpreted with evolutionary models, implies a mass of $\sim$0.3~$M_{\mathrm{Jup}}$. This detection was independently confirmed using the Near-Infrared Camera (NIRCam), which further constrained the photometric properties of the planet and the detailed morphology of the disk in scattered light \citep{Crotts+2025}. The planet appears to reside within a dust-depleted region of the disk, consistent with a scenario in which it gravitationally shapes the outer disk structure.

High-angular-resolution observations with the Very Large Telescope (VLT) using the Spectro-Polarimetric High-contrast Exoplanet REsearch (SPHERE) revealed three main belts at approximately 7, 25, and 52~au together with clumps and extended asymmetric features \citep{Olofsson+2018}. The Atacama Large Millimeter/submillimeter Array (ALMA) further resolved the disk from a nearby galaxy, effectively ruling out background contamination \citep{Bayo+2019}. Multiple-belt debris-disk architectures are frequently interpreted as the dynamical signature of unseen planets sculpting gaps between spatially separated planetesimal populations. In particular, \citet{Lazzoni+2018} applied chaotic-zone scaling relations to debris disks to relate gap widths to the mass and multiplicity of perturbing planets. When applying this methodology to TWA~7, \citet{Olofsson+2018} suggested that reproducing the entire 7-25~au cavity likely requires multiple sub-Jovian companions.

While the presence of a gap is commonly interpreted as evidence of a planetary companion, finer structures in the disk can provide additional dynamical constraints. For instance, a scattered-light feature aligned with the projected orbit of TWA~7~b has recently been reported \citep{Lagrange+2025}. Its morphology appears horseshoe-like, which is potentially consistent with a co-orbital configuration. Co-orbital architectures are known to arise only under specific dynamical conditions, with their long-term stability depending sensitively on the planet-star mass ratio, orbital eccentricity, and external perturbations \citep[e.g.,][]{Dermott+1981, Cuk+2012, Leleu+2018}. The Trojan population of Jupiter provides the most prominent example of such long-lived configurations \citep{Marzari+2002}. However, beyond the Solar System, co-orbital systems remain observationally elusive, with only a few tentative candidates reported so far \citep[e.g.,][]{Leleu+2019, Balsalobre-Ruza+2023}. If confirmed, the presence of a co-orbital structure in TWA~7 would therefore provide both a rare observational case and a sensitive probe of the dynamical processes shaping young planetary systems, offering insight into how such fragile configurations can emerge and persist in a disk environment.

Furthermore, observations of the inner disk have revealed a sharp decline in scattered light near 23~au \citep{Ren+2021}, marking a well-defined inner edge. Given its large separation at $\sim$52~au, TWA~7~b is unlikely to account for this inner truncation on its own. Such a feature therefore points to the possible presence of an additional yet undetected inner companion, similar to the case of $\beta$~Pictoris, where dynamical modeling has constrained an additional planet shaping the inner cavity \citep{Lacquement+2025}.

Altogether, TWA 7 provides an exceptional laboratory to investigate the dynamical interplay between planets and debris disks, including possible co-orbital configurations and multi-belt architectures. In this study, we adopt a purely gravitational framework and combine observational constraints with numerical N-body simulations and secular theory to explore the architecture of this young planetary system. We focus specifically on the sharp inner edge of the central belt at 23~au and do not attempt to model the full extent of the cavity between the 7 and 25~au belts. In Sect.~\ref{twa7b_constraints_coorbital}, we first revisit the properties of the outer planet TWA~7~b, whose putative co-orbital material offer a unique dynamical constraint on its eccentricity. The stability of this population requires that TWA~7~b remains on a nearly circular orbit, a condition adopted as a reference for the subsequent analysis. In Sect.~\ref{twa7c_constraints_inner_edge}, we use high-contrast imaging limits together with numerical simulations of the debris disk to relate the morphology of its inner edge to the possible presence of an unseen inner planet. In Sect.~\ref{twa7c_constraints_secular_coupling}, we investigate the secular coupling between TWA~7~b and this hypothetical inner companion to derive additional constraints on its eccentricity and dynamical compatibility with the observed system. Altogether, these complementary approaches provide a coherent framework to delineate the parameter space where an additional planet could remain undetected yet dynamically significant.

\section{Co-orbital constraint on TWA~7~b}
\label{twa7b_constraints_coorbital}

Recently, TWA~7~b was detected with JWST/MIRI at a projected separation of $\sim$52~au, and it has been estimated to have a mass of $\sim$0.3~$M_{\mathrm{Jup}}$. In addition to this direct detection, the disk surrounding TWA~7 exhibits an extended horseshoe-shaped feature, which could correspond to co-orbital material librating around the planetary orbit. Owing to the long orbital period at such a large separation, direct astrometric constraints on the orbit are unlikely to be obtained in the foreseeable future. The disk morphology therefore offers a valuable indirect probe of the orbital properties of the planet. In this section, we exploit the characteristics of horseshoe dynamics to quantify these constraints, with a particular emphasis on the orbital eccentricity.

\subsection{Definition and context}
\label{twa7b_constraints_coorbital_definition}

Co-orbital configurations refer to bodies trapped in a 1:1 mean-motion resonance with a planet and are classically described within the framework of the restricted three-body problem \citep{Murray+1999}. These configurations are organized around the triangular Lagrange points L$_4$ and L$_5$ and the collinear point L$_3$.

Two principal dynamical classes are usually distinguished. In tadpole orbits, objects librate with moderate amplitudes around either L$_4$ or L$_5$, forming Trojan swarms. In contrast, horseshoe orbits occupy a broader region of phase space, with trajectories that alternately lead and trail the planet and encompass L$_3$, L$_4$, and L$_5$ in the rotating frame.

Hydrodynamical simulations have shown that co-orbital planetary configurations can naturally arise during the gas-rich phase of planet formation as a result of convergent migration and resonance trapping within protoplanetary disks \citep[e.g.,][]{Laughlin+2002,Cresswell+2008,Pierens+2014}. In particular, \citet{Cresswell+2008} reported that more than 30\% of their multi-planet simulations resulted in the spontaneous formation of co-orbital pairs, as migrating embryos became trapped near the Lagrange points through phase-protection mechanisms. These systems tend to remain dynamically cold, with low eccentricities and inclinations maintained by gas-induced damping, which suppresses orbital excitation and supports co-orbital stability.

\subsection{Stability and eccentricity threshold}
\label{twa7b_constraints_coorbital_stability}

After gas dispersal, the evolution of co-orbital configurations is largely governed by gravitational dynamics. In this regime, horseshoe orbits are intrinsically fragile, as their libration domain encompasses the unstable Lagrange point $L_3$ and is structured by a dense network of secondary resonances. These resonances fragment the horseshoe phase-space into narrow chaotic layers and stable islands, leading to slow chaotic diffusion even in dynamically cold systems \citep[e.g.,][]{Robutel+2013, Leleu+2015, Paez+2015}.

Despite this intrinsic chaotic structure, horseshoe configurations can remain stable in practice if a geometric protection mechanism operates. Using long-term numerical integrations, \citet{Cuk+2012} showed that the effective stability of horseshoe orbits is primarily controlled by the minimum distance reached between the co-orbital body and the planet during the libration cycle. As long as this distance remains larger than a few Hill radii, close encounters are avoided, and the horseshoe motion can persist over millions of orbital periods. In circular configurations, this geometric protection is typically achieved for planet-to-star mass ratios bellow $\mu_{crit} \simeq 8.3 \times 10^{-4}$.

In the case of TWA~7~b, the inferred planet-to-star mass ratio is $\mu_b \simeq 6.2 \times 10^{-4}$, thus safely below the critical threshold. This implies that in the absence of additional perturbations and for a nearly circular orbit, horseshoe configurations around TWA~7~b are not intrinsically unstable and can survive for timescales well in excess of the age of the system.

However, this geometric protection is highly sensitive to planetary eccentricity. A finite eccentricity renders the co-orbital potential explicitly time dependent, broadens secondary resonances, and enhances resonance overlap within the horseshoe region. As a result, chaotic diffusion becomes significantly more efficient and rapidly drives horseshoe trajectories toward destabilizing close encounters. Numerical simulations by \citet{Leleu+2018} have shown that even modest planetary eccentricities strongly reduce the effective stability domain of horseshoe motion. In particular, for eccentricities of order $e \simeq 0.05$, the critical mass ratio for long-term stability drops to values of $\mu_{crit} \simeq 3 \times 10^{-4}$, which falls just below the value for TWA~7~b.

Our own simulations are consistent with this picture. We find that a stable horseshoe motion around TWA~7~b is maintained for planetary eccentricities up to $e_b \simeq 0.05$, while higher eccentricities lead to rapid erosion of the co-orbital region. This identifies a clear dynamical constraint: Although the low mass ratio of TWA~7~b satisfies the necessary geometric condition for horseshoe stability, the persistence of a horseshoe-like population requires the planet to remain on a nearly circular orbit. The existence of such a structure therefore places a stringent upper limit of $e_b \lesssim 0.05$ on the eccentricity of TWA~7~b, a constraint significantly stronger than those currently accessible through direct imaging alone. This bound is adopted in the following sections as a key dynamical condition when assessing the compatibility of additional planetary companions with both the observed disk architecture and the long-term survival of the co-orbital material.

\section{Constraint from the inner edge of the disk}
\label{twa7c_constraints_inner_edge}

The sharply defined inner edge of the debris disk, observed near 23~au, cannot be solely attributed to the gravitational influence of TWA~7~b, located at $\sim$52~au. As established in Sect.~\ref{twa7b_constraints_coorbital}, the persistence of a stable co-orbital population around TWA~7~b indicates that this outer planet follows a nearly circular orbit, implying that the sculpting of the inner edge of the disk must instead result from an additional currently undetected inner companion. We refer to this hypothetical planet as TWA~7~c. In this section, we combine observational detection limits and numerical simulations of the debris disk to constrain the possible mass and position of TWA~7~c from the morphology of the inner edge before investigating how its dynamical coupling with TWA~7~b further restricts its eccentricity in Sect.~\ref{twa7c_constraints_secular_coupling}.

\subsection{Radial and mass domain considered}
\label{twa7c_constraints_inner_edge_constraints}

High-contrast VLT/SPHERE observations of TWA~7 probe the Jovian-mass regime at separations comparable to the disk inner edge, and no additional companion was detected \citep{Olofsson+2018}. Planets with masses on the order of $1\,M_{\rm Jup}$ or larger in this region are therefore strongly disfavored.

At the same time, a planet capable of truncating the disk at 23~au must orbit sufficiently close to this boundary to dynamically shape it. In the framework of resonance overlap, the radial extent of the chaotic region carved by a planet is primarily controlled by its mass and eccentricity \citep{Wisdom1980, Mustill+2012, Morrison+2015}. For moderate eccentricities and sub-Jovian masses, this cleared region remains confined to a limited radial range, on order of ten astronomical units. A perturber located interior to the disk edge by more than roughly this distance would therefore be unable to reproduce the observed sharp truncation.

Taken together, these arguments restrict the relevant parameter space to $13 \leq a_{c} \leq 23$~au and $m_{c} \leq 1\,M_{\rm Jup}$, where 13~au coincides with the SPHERE inner working angle (IWA). This domain defines the scope of the present study.

\subsection{Numerical modeling of the inner edge}
\label{twa7c_constraints_inner_edge_method}

We present a series of dynamical simulations designed to test whether an additional planet interior to TWA~7~b could explain the sharp inner edge of the debris disk observed at $\sim$23~au. The methodology adopted here closely follows that developed in our previous study of disk sculpting dynamics in the $\beta$~Pictoris system, where an additional planet was invoked to reproduce the sharp inner edge \citep{Lacquement+2025}. This approach has been proven to be robust and reproducible for constraining planetary architectures based on disk morphology.

The modeled system consists of the central star, the outer planet TWA~7~b, and a hypothetical inner planet, TWA~7~c. The orbital parameters of TWA~7~b were fixed according to the constraints derived in Sect.~\ref{twa7b_constraints_coorbital}, with $m_b \simeq 0.3\,M_{\mathrm{Jup}}$, $a_b \simeq 52$~au, and $e_b \lesssim 0.05$. Since TWA~7~c is assumed to be part of the same planetary system, we adopted a coplanar configuration. The other angular parameters, including the longitude of the ascending node, argument of periastron, and mean longitude, were randomly selected, as these quantities undergo secular precession under the gravitational influence of TWA~7~b and have only a minor long-term dynamical impact. The key free parameters for TWA~7~c are its mass ($m_c$), semi-major axis ($a_c$), and eccentricity ($e_c$), which are systematically explored over a regular grid.

Each simulation includes 400,000 massless test particles representing planetesimals, initially distributed between 20 and 100~au. The particles are non-interacting and evolve under the gravitational potential of the star and the two planets. Initial eccentricities were drawn uniformly between 0 and 0.05, and inclinations were drawn between 0$^{\circ}$ and 2$^{\circ}$ relative to the invariant plane of the system. The other angular elements were randomly distributed between 0$^{\circ}$ and 360$^{\circ}$ to ensure an unbiased sampling of the phase space. During the integrations, particles were removed if they collided with the central star (distance smaller than 0.005~au) or if they escaped beyond 200~au. Likewise, unbound particles were discarded once they moved sufficiently far from the system. In the two-planet configurations explored here, approximately 20\% of the initial test particles were removed over 10~Myr, depending on the mass and orbital parameters of the inner companion. All results presented in this work were computed using only the particles that remained in the simulation at the end of the integrations.

Integrations were performed using the regularized mixed variable symplectic integrator, version 3 \citep[RMVS3;][]{Levison+1994}, derived from the Wisdom--Holman scheme \citep{Wisdom+1991}, and we optimized it for long-term evolution with potential close encounters. Each simulation spanned 10~Myr, consistent with the estimated age of the system.

To enable direct comparison with VLT/SPHERE observations, we accounted for its angular resolution in the H~band ($\lambda=1.6\,\mu$m), corresponding to a diffraction limit of $\sim$49~mas, or $\sim$1.7~au at the distance of TWA~7. The simulated surface density maps were convolved with a Gaussian kernel with a full width at half maximum equal to this resolution to mimic the effective point spread function of the instrument. From the convolved maps, we extracted the azimuthally averaged radial surface density profile. 

The location of the inner edge of the disk was defined as the point of the steepest positive gradient in surface density corresponding to the maximum of the first derivative. This position was refined by fitting a Gaussian to the derivative peak, yielding precise and reproducible measurements of both the edge position and the transition width. By comparing the simulated and observed edge positions, we identified the combinations of $(m_c, a_c, e_c)$ that reproduce the observed value of 23~au.

\subsection{Results for circular orbits}
\label{twa7c_constraints_inner_edge_results}

\begin{figure*}[ht!]
\makebox[\textwidth]{
\includegraphics[width=0.3333\textwidth]{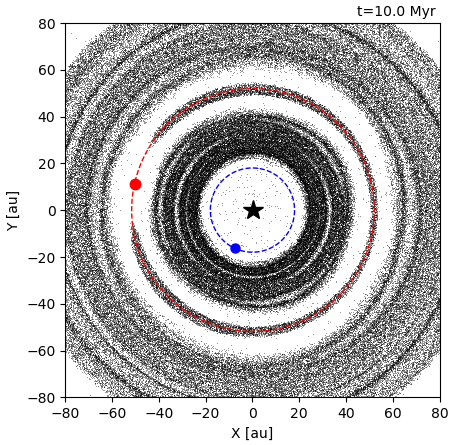} \hfil
\includegraphics[width=0.3315\textwidth]{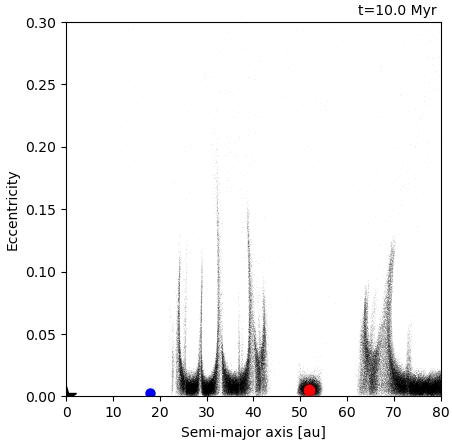} \hfil
\includegraphics[width=0.321\textwidth]{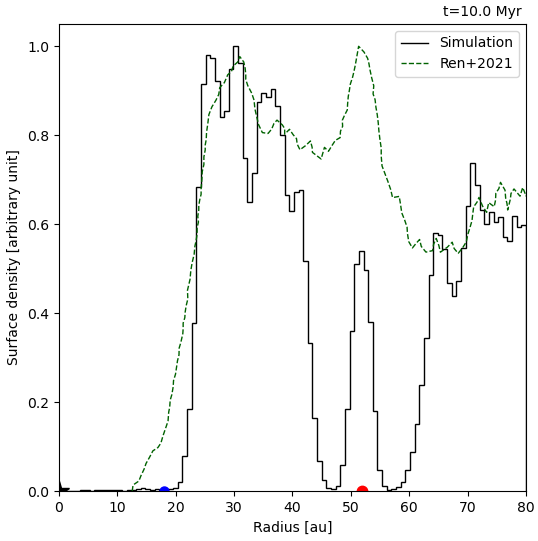}}
\caption{Dynamical status at the end of an example of a simulation of the dynamics of the TWA~7 planetary system with two planets, TWA~7~b (red), an additional planet TWA~7~c (blue), and the disk of planetesimals. The initial orbital parameters of TWA~7~b are $m_b=0.3$~\mjup, $a_b=52$~au \citep{Lagrange+2025}, and $e_b=0$, consistent with the constraints derived in Sect.~\ref{twa7b_constraints_coorbital}. In this example, the initial orbital parameters of TWA~7~c are $m_c=0.4$~\mjup, $a_c=18$~au, and $e_c=0$. \emph{Left:} View of the system from above. The planetesimals are depicted with small black dots, and the orbits of the planets are indicated with dashed colored lines. \emph{Middle:} View of the system in terms of semi-major axis and eccentricity. \emph{Right:} Radial profile of the planetesimal disk (solid black line) superimposed to the observations of  \cite{Ren+2021} (green dotted line). This configuration reproduces the observed inner edge of the disk while preserving a population of horseshoe co-orbitals along the orbit of TWA~7~b.}
\label{fig:ExSimuValide}
\end{figure*}

\begin{figure}[]
\centering
\includegraphics[width=\columnwidth]{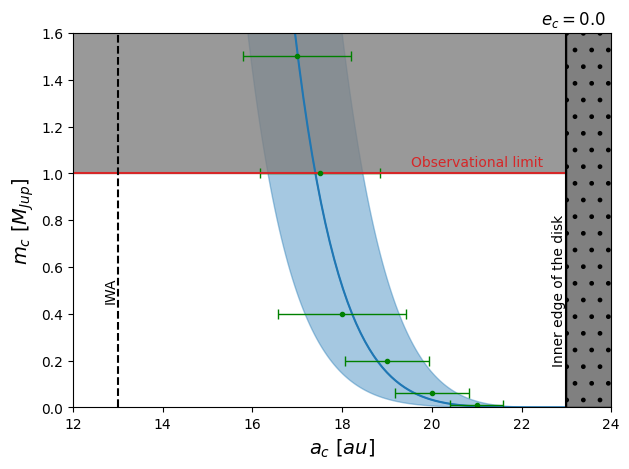}
\caption{Combinations of the mass and initial semi-major axis of TWA~7~c within the observational limits that successfully reproduce the inner edge of the disk at 23~au. For several values of $m_c$, the corresponding acceptable ranges of $a_c$ are shown in green. These ranges reflect the width of the inner edge of the disk as defined in Sect.~\ref{twa7c_constraints_inner_edge_method}. A power-law fit is overlaid in blue, accounting for the extent of each semi-major-axis range. Shaded areas correspond to regions that are not dynamically accessible to TWA~7~c.}
\label{fig:simus_c_e000}
\end{figure}

We first considered circular orbits ($e_c = 0$) to isolate the influence of the mass and semi-major axis of the planet. Figure~\ref{fig:ExSimuValide} shows a representative simulation reproducing the observed disk edge at 23~au, illustrating the corresponding disk morphology and radial surface density profile. The overall set of simulations for circular orbits, summarized in Fig.~\ref{fig:simus_c_e000}, reveals a clear and systematic trend: Lower-mass planets must orbit closer to the disk edge to reproduce the observed truncation, whereas higher-mass planets can carve the same feature from larger distances. The relation between $m_c$ and $a_c$ is well described by a power-law fit consistent with theoretical predictions for the chaotic zone width of \citet{Wisdom1980} and \citet{Mustill+2012}.

The strong agreement between analytical expectations and numerical results supports the interpretation that the inner edge of the disk is shaped by gravitational clearing from an interior planet. Having established how the mass and semi-major axis of the planet control the location of the inner edge of the disk in the circular case, we then investigated how a finite eccentricity of TWA~7~c could modify this configuration. To this end, we made use of secular coupling with TWA~7~b, which provides an indirect yet powerful means of constraining the eccentricity of the inner planet.

\section{Constraint from secular coupling}
\label{twa7c_constraints_secular_coupling}

The presence of co-orbital material associated with TWA~7~b implies that this planet follows a nearly circular orbit, with an eccentricity constrained to $e_b \lesssim 0.05$, defined in Sect.~\ref{twa7b_constraints_coorbital}. However, secular interactions with an inner companion can gradually excite its eccentricity through long-term angular momentum exchange \citep[e.g.,][]{Kane+2014, Read+2016}. Preserving the observed dynamical coldness of TWA~7~b therefore imposes indirect but stringent limits on the properties of TWA~7~c. In this section, we apply secular theory to quantify the coupling between the two planets and identify the region of parameter space for which the eccentricity of TWA~7~b remains consistent with the existence of a stable co-orbital population.

\subsection{Secular excitation in the low-eccentricity regime}
\label{twa7c_constraints_secular_coupling_theory}

For systems that are nearly circular and coplanar, secular interactions can be accurately described using the linear Laplace-Lagrange theory \citep{Murray+1999}. Within this framework, the complex eccentricity vectors $z_j = e_j \exp(i \varpi_j)$, where $\varpi_j$ is the longitude of periastron,  evolve according to
\begin{equation}
    \frac{d\vec{z}}{dt} = i\mathbf{A} \cdot \vec{z},
\end{equation}
where $\mathbf{A}$ is the secular interaction matrix whose coefficients depend on the planetary masses and semi-major axes through Laplace coefficients. These coefficients quantify the strength of the mutual gravitational coupling between the two orbits and are typically evaluated numerically or obtained from standard tabulations (see Appendix~B of \citealt{Murray+1999}).

As we have two planets, the general solution can be written as a superposition of two eigenmodes, each with its own frequency ($g_j$) and eigenvector ($v_j$):
\begin{equation}
    \vec{z}(t) = C_1 e^{i g_1 t} \vec{v_1} + C_2 e^{i g_2 t} \vec{v_2},
\end{equation}
with amplitudes $C_1$ and $C_2$ determined by the initial conditions. This describes the secular exchange of eccentricity between the two planets over long timescales.

Following the formalism of \citet{Read+2016} but applying it to an inverted symmetric configuration, we considered an initially circular outer planet ($e_b(0)=0$) and a low-eccentricity inner planet ($e_c(0)\neq0$). It can be shown that in the linear regime, the maximum eccentricity reached by the outer planet due to secular coupling with the inner one is then given by
\begin{equation}
\max \left[e_b\right] = e_c(0) \left( \frac{L_b}{L_c} + \frac{1}{4} f^2 \left( \frac{L_b}{L_c} - 1 \right)^2 \right)^{-\frac{1}{2}},
\label{eq:eb_max}
\end{equation}
where
\begin{itemize}
  \item \( L_j = m_j \sqrt{a_j} \) is the angular momentum proxy of planet \( j \),
  \item \( f = \frac{b_{3/2}^{(1)}(\alpha)}{b_{3/2}^{(2)}(\alpha)} \) is the ratio of Laplace coefficients, and
  \item \( \alpha = \min \left[\frac{a_b}{a_c}, \frac{a_c}{a_b} \right] \).
\end{itemize}
This expression captures how the secular excitation of TWA~7~b depends on the mass, orbital spacing, and eccentricity of TWA~7~c.

\subsection{Numerical evaluation of the coupling effect}
\label{twa7c_constraints_secular_coupling_method}

We showed in Sect.~\ref{twa7c_constraints_inner_edge} that a planet on a circular orbit can reproduce the observed inner edge of the debris disk at 23~au. We now extend this analysis to assess how a finite eccentricity of the hypothetical planet TWA~7~c could modify both the disk morphology and the dynamical stability of the system. Specifically, we aim to determine the range of $(m_c, a_c, e_c)$ combinations that simultaneously preserve the observed inner edge and maintain the stability of co-orbitals around TWA~7~b.

To explore this efficiently without repeating the full suite of disk simulations, we adopted a simplified approach that preserves the location of the inner edge of the disk. For each tested value of $e_c$, we identified the corresponding $(m_c, a_c)$ combinations that maintain the same apastron distance as the circular reference case. This ensures that the planet continues to interact with the same region of the disk, thereby keeping the truncation radius near 23~au.

In a second step, we evaluated whether such eccentric configurations remain dynamically compatible with the stability of TWA~7~b and its co-orbital population. For each $(m_c, a_c, e_c)$ configuration, we computed the eccentricity induced on TWA~7~b using the analytical expression Eq.~\ref{eq:eb_max}. This allowed us to map the secular excitation of TWA~7~b across the $(m_c, a_c)$ parameter space for different eccentricities of TWA~7~c. We then compared the resulting $e_b$ with the upper limit of 0.05 required for the long-term stability of horseshoe-like co-orbitals. Configurations that exceeded this threshold were considered dynamically inconsistent and excluded from the allowed parameter space.

\subsection{Results for the system architecture}
\label{twa7c_constraints_secular_coupling_results}

\begin{figure*}[]
    \makebox[\textwidth]{
    \includegraphics[width=0.3333\textwidth]{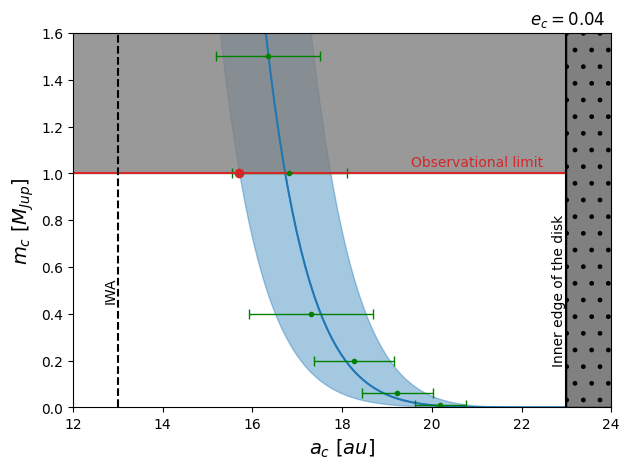} 
    \includegraphics[width=0.3333\textwidth]{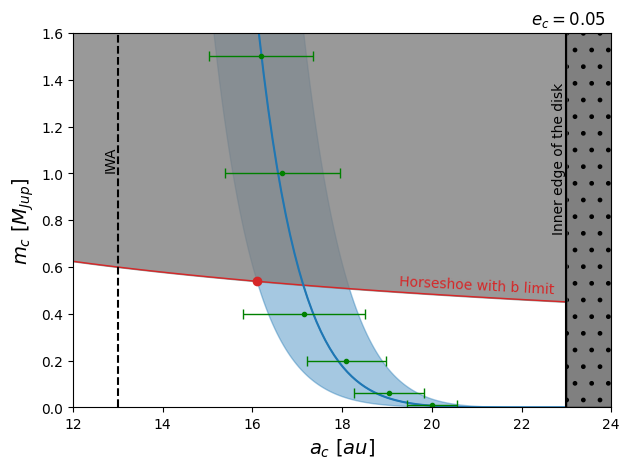}
    \includegraphics[width=0.3333\textwidth]{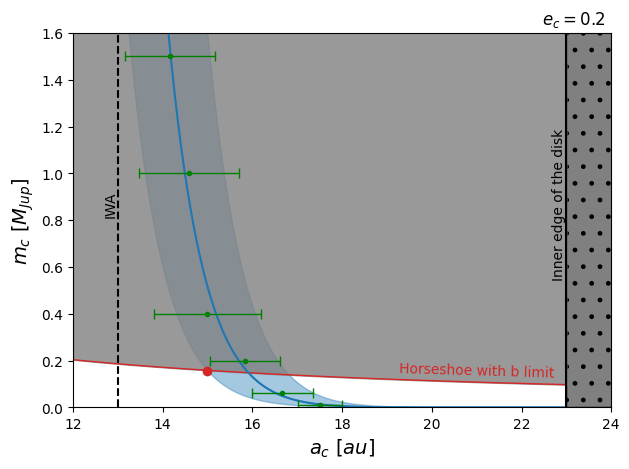}}
    \caption{Allowed configurations for a possible inner planet in the $(a_c, m_c)$ plane for increasing values of initial eccentricity, $e_c$, under the constraint that the simulated disk reproduces the observed inner edge at 23~au. The plotting conventions are the same as in Fig.~\ref{fig:simus_c_e000}. The red line marks the upper mass limit set by observations and, when applicable, by the additional requirement that the outer planet TWA~7~b remains dynamically cold enough to support stable horseshoe-like co-orbitals (Sect.~\ref{twa7b_constraints_coorbital}). The red dot indicates the intersection point where, for a given eccentricity, all solutions above this limit are excluded. As $e_c$ increases, this constraint becomes increasingly restrictive, eventually ruling out all configurations compatible with the observed disk morphology.}
    \label{fig:twa7c_e}
\end{figure*}

\begin{figure}
    \centering
    \includegraphics[width=\columnwidth]{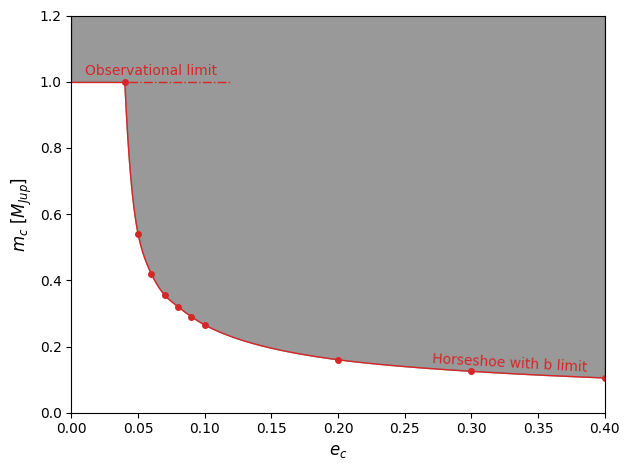}
    \caption{Accessible mass of the additional planet TWA~7~c as a function of its eccentricity $e_{c}$, based on the results of Fig.~\ref{fig:twa7c_e}, whose plot conventions are similar.}
    \label{fig:Bpic-d-constraint-em}
\end{figure}

\begin{figure}[]
    \centering
    \includegraphics[width=\columnwidth]{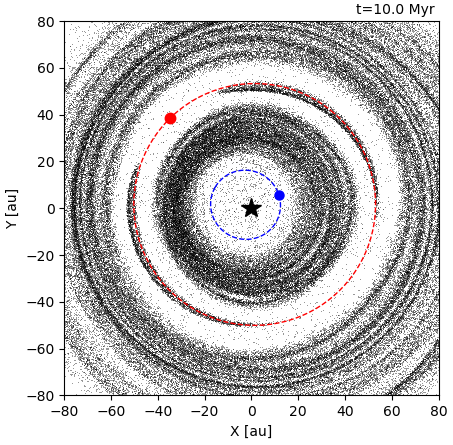}
    \caption{View from above of the same simulation as in Fig.~\ref{fig:ExSimuValide}, but with an initial eccentricity of $e_c = 0.2$ for the additional planet TWA~7~c. This configuration reproduces the observed inner edge of the disk but fails to preserve a stable population of horseshoe co-orbitals along the orbit of TWA~7~b, leaving only two separate arcs consisting of particles moving on tadpole orbits.}
    \label{fig:ExSimuInValide}
\end{figure}

Some resulting maps are shown in Fig.~\ref{fig:twa7c_e}, where each panel corresponds to a different value of \( e_c \). The red lines indicate the upper mass limit compatible with the observational limit and horseshoe stability. The results reveal a clear trend: As \( e_c \) increases, the same disk edge can be reproduced by planets on more eccentric orbits but at the cost of stronger secular forcing on TWA~7~b. For \( e_c \lesssim 0.04 \), the constraint from co-orbital stability remains weaker than the direct imaging limits. Beyond this threshold, the secular perturbations from TWA~7~b become dominant, significantly limiting the permissible mass of TWA~7~c. As a result, the allowed mass of TWA~7~c drops sharply, ultimately ruling out nearly all viable configurations at higher eccentricities. This trend is summarized in Fig.~\ref{fig:Bpic-d-constraint-em}, which displays the maximum permitted mass of TWA~7~c as a function of its eccentricity. One such excluded scenario is illustrated in Fig.~\ref{fig:ExSimuInValide}, where no co-orbital particles remain near the $L_3$ point of TWA~7~b. While some co-orbitals on tadpole orbits around the $L_4$ and $L_5$ points still survive, the lack of a continuous horseshoe-shaped ring renders this configuration inconsistent with observations.

These results demonstrate that maintaining both the observed disk morphology and the persistence of a co-orbital population requires that both TWA~7~b and TWA~7~c remain on low-eccentricity orbits. Any significant excitation of either planet would destabilize the horseshoe region and erase the co-orbital material, thereby breaking the delicate dynamical balance observed in the system.

\section{Conclusion and discussion}
\label{Conclusion}

In this study, we have investigated the architecture of the TWA~7 planetary system through the morphology of its debris disk. The possible detection of a horseshoe co-orbital structure associated with TWA~7~b provides a strong dynamical constraint: The long-term survival of such a configuration requires the planet to maintain a low eccentricity, $e_b \lesssim 0.05$. Within the Laplace-Lagrange secular framework, this condition propagates inward and restricts the admissible properties of any additional companion, confining viable architectures to low-mass planets on nearly circular orbits. These combined constraints indicate that TWA~7 has experienced only limited dynamical excitation.

Within this framework, the outer planet TWA 7 b, detected at $\sim$52~au, cannot by gravitational interactions alone account for the sharp inner edge observed at $\sim$23~au. Our $N$-body simulations instead indicate that an undetected inner companion provides a natural explanation for this truncation. By combining dynamical modeling with current direct-imaging limits, we identified a narrow region of parameter space in which a sub-Jovian planet orbiting between 13 and 23~au on a low-eccentricity orbit simultaneously reproduces the observed disk morphology and remains consistent with observational constraints.

We emphasize that in the present study, we focused on the origin of the truncation at $\sim$23~au, and our aim was not to reproduce the full radial extent of the 7-25~au cavity reported by \citet{Olofsson+2018}, who showed that clearing such a wide gap likely requires multiple planets. Within our framework, a single planet orbiting between 13 and 23~au can reproduce the sharp outer edge at 23~au, but it cannot by itself account for the entire 7-25~au depletion. Our results are therefore consistent with a multiplanet architecture, but they also place tighter dynamical constraints through the additional requirement imposed by the co-orbital structure.

More broadly, recent analyses of Gaia data combined with radial velocity and direct imaging suggest the presence of an additional inner planet at separations on the order of 2-3~au, possibly with a mass of several Jupiter masses \citep{Lagrange+2026}. Although such a companion cannot, through gravitational interactions alone, account for the truncation at 23~au, it may contribute to shaping the innermost regions of the system and participate in a coupled architecture. Preliminary numerical experiments indicate that its dynamical impact depends sensitively on its mass: While a $\sim$7 $M_{\rm Jup}$ planet excites the disk and disrupts the horseshoe co-orbitals, a companion on the order of $\sim$2 $M_{\rm Jup}$ remains compatible with both the survival of the co-orbital population and the observed disk structure. A comprehensive exploration of such coupled configurations would require a dedicated survey of the parameter space, but this lies beyond the scope of the present study.

To connect our dynamical model with the observed disk morphology, we assumed that the dust distribution broadly traces the underlying parent planetesimal population. Radiation pressure increases grain eccentricities and redistributes dust outward while preserving their pericenters near the production site \citep[e.g.,][]{Wyatt2006,Augereau+2006}. Both the inner edge at 23~au and the horseshoe structure around TWA~7~b therefore correspond to genuine concentrations of parent planetesimals. In contrast, as illustrated in Fig.~\ref{fig:ExSimuValide}, the dust distribution extends beyond the more sharply truncated parent planetesimal population. This broader morphology may result from eccentric grains released from these regions, which would transiently populate the gap around TWA~7~b.

Despite this consistency between the dynamical model and the observed structures, our numerical approach relies on a purely gravitational N-body framework and neglects additional physical processes that may influence the long-term evolution of the system. Given the young age of TWA~7, residual gas may still be present \citep{WilliamsCieza2011} and could damp orbital excitation \citep{Tanaka+2004}, potentially enhancing the stability of co-orbital material in mildly perturbed configurations. Mean-motion resonances may also play a role in shaping the system architecture. For a companion located between 13 and 23~au, low-order resonances with TWA~7~b, such as the 4:1 at $\sim$21~au, the 5:1 at $\sim$18~au, and the 6:1 at $\sim$16~au, are naturally expected. These resonances could contribute to eccentricity excitation, temporary resonant trapping, or additional dynamical structuring of the disk \citep{Murray+1999}. Moreover, horseshoe co-orbitals are themselves sensitive to internal secondary resonances, which may either stabilize or destabilize trajectories through chaotic diffusion \citep{Robutel+2013}. Finally, our secular analysis is formulated within the framework of linear Laplace-Lagrange theory, which assumes low eccentricities and coplanarity and neglects higher-order nonlinear coupling terms. This approximation is consistent with the dynamically cold configurations required by the co-orbital constraint. However, extending the analysis to include gas dynamics, resonant interactions, and nonlinear secular effects would provide a more comprehensive description of the system and help assess the robustness of our conclusions.

This work contributes to ongoing efforts to use debris-disk morphology to infer planetary architectures, as highlighted by broader surveys such as \citet{Pearce+2022}, who compiled numerous systems exhibiting structures likely shaped by unseen planets. In this context, TWA~7 stands out because the possible co-orbital structure provides an additional and unusually stringent constraint on the mass and eccentricity of unseen companions. By combining disk morphology with co-orbital dynamics, it offers a particularly powerful diagnostic of the dynamical state of the system. The secular framework developed here provides an efficient and flexible tool for exploring such constraints and can be readily applied to other young debris disks with similarly structured architectures.
 
\begin{acknowledgements} 
We are grateful to an anonymous referee for feedback that helped us improve this manuscript. All computations presented in this paper were performed using the GRICAD infrastructure (\texttt{https://gricad.univ-grenoble-alpes.fr}), which is supported by Grenoble research communities. 
\end{acknowledgements}

\bibliographystyle{aa.bst}
\bibliography{bibli}

\end{document}